\documentclass{ws-procs9x6}
\pdfoutput=1
\usepackage[colorlinks=true,urlcolor=blue]{hyperref}

\def\urltilda{\kern -.15em\lower .7ex\hbox{\~{}}\kern .04em}
\begin{document}

\title{THE DARKSTARS CODE: A PUBLICLY AVAILABLE DARK STELLAR EVOLUTION PACKAGE}

\author{PAT SCOTT$^*$ and JOAKIM EDJS\"O}
\address{Oskar Klein Centre for Cosmoparticle Physics \&\\
Department of Physics, Stockholm University,\\AlbaNova University Centre, SE-106 91 Stockholm, Sweden\\
$^*$Presenting author; pat@fysik.su.se}

\author{MALCOLM FAIRBAIRN}
\address{PH-TH, CERN, Geneva, Switzerland \& King's College London, WC2R 2LS, UK}

\begin{abstract}
We announce the public release of the `dark' stellar evolution code \textsf{DarkStars}.  The code simultaneously solves the equations of WIMP capture and annihilation in a star with those of stellar evolution assuming approximate hydrostatic equilibrium.  \textsf{DarkStars} includes the most extensive WIMP microphysics of any dark evolution code to date.  The code employs detailed treatments of the capture process from a range of WIMP velocity distributions, as well as composite WIMP distribution and conductive energy transport schemes based on the WIMP mean-free path in the star.  We give a brief description of the input physics and practical usage of the code, as well as examples of its application to dark stars at the Galactic centre.
\end{abstract}

\keywords{cosmology; stellar evolution; dark matter; WIMPs; galactic centre; dark stars}

\bodymatter

\section*{}
The last two years have seen strong interest in the impacts of dark matter upon stellar structure and evolution.  The predominant focus has been on self-annihilating WIMP (weakly-interacting massive particle) dark matter, because it has the ability to affect stellar structure by annihilating in stellar cores \cite{Moskalenko07, Bertone07, Fairbairn08, Iocco08a, Yoon08, Taoso08, Scott09} and collapsing protostellar halos \cite{Spolyar08, Freese08c, Iocco08b, Ripamonti09, Spolyar09}.  Interest has been driven by the prospect of providing constraints upon the nature of dark matter\cite{Bertone07,Scott09}, by intrinsic curiosity in the resultant `dark stars' themselves\cite{Iocco08a, Freese08a, Spolyar09}, and by their possible impacts upon early-universe processes like reionisation \cite{Natarajan08a, Schleicher08}.

We have previously discussed the possibility that main-sequence dark stars could exist at the centre of our own Galaxy \cite{Fairbairn08,Scott08a,Scott08b,Scott09}.  In those papers we utilised a form of the standard stellar evolution code \textsc{stars} \cite{Eggleton71,Eggleton72,Pols95} modified to include the effects of dark matter capture and annihilation.  This modified code is \textsf{DarkStars}, and in these proceedings we announce its public release.  \textsf{DarkStars} is written in Fortran95, and can be freely downloaded from \href{http://www.fysik.su.se/~pat/darkstars}{\tt http://www.fysik.su.se/{\urltilda}pat/darkstars}.  Below we give outlines of the code's input physics and practical usage, along with some simple examples of stars evolved with it.

\textsf{DarkStars} includes gravitational capture of WIMPs from the galactic halo via the full equations of Gould\cite{Gould87b}, including both spin-dependent and spin-independent scattering on the 22 most important atomic nuclei.  The capture routines are adapted from the solar capture code in \textsf{DarkSUSY}\cite{darksusy}.  Capture can be performed semi-analytically from either a standard isothermal WIMP halo or an isothermal halo where the WIMP velocity distribution has been truncated at the local escape velocity.  Alternatively, numerical capture calculations can be performed on a velocity distribution derived\cite{Fairbairn09} from the Via Lactea \cite{vialactea} simulation of a Milky Way-type galaxy, or any other arbitrary, user-supplied velocity distribution.  

The distribution of WIMPs with height in a star is obtained by interpolating between two limiting distributions according to the value of the WIMP mean-free path in the star: one corresponding to WIMPs with very long mean-free paths, the other to WIMPs with very short mean-free paths.  Conductive energy transport by weak-scattering events between atomic nuclei and WIMPs is taken into account in a manner consistent with this distribution: the conductive luminosity at each height is approximated by rescaling the known expression for the conductive luminosity at short mean-free paths, according to the actual value of the mean free path in the star.  The annihilation luminosity at each height in the star is simply calculated as the product of the annihilation cross-section and square of the local WIMP density, and fed along with the conductive luminosity into the luminosity equation in the stellar solver.  Full technical details of the input physics for \textsf{DarkStars} can be found in Ref.~\refcite{Scott09}.

\textsf{DarkStars} operates with a simple text-file input, containing a series of switches and physical parameters with which to perform a particular evolutionary run.  Switches allow choices between analytical and numerical capture, different halo velocity distributions, the inclusion or exclusion of annihilation and conductive energy transport effects, and the option to run in a special `reconvergence mode' where the solution obtained at each timestep is converged twice (see Ref.~\refcite{Scott09} for details).  Runs can be saved and restarted at will, and the input format provides the ability to make periodic saves during the course of a single evolutionary run.

The user can specify the WIMP mass, spin-dependent, spin-independent and annihilation cross-sections, as well as the stellar mass and metallicity, the initial population of WIMPs in the star and the ultimate percentage of energy lost to neutrinos in each annihilation.  One may also opt to specify a constant stellar velocity through a WIMP halo with some particular local density and velocity dispersion, located at a position with a single well-defined Galactic escape velocity.  Alternatively, runs can be performed along user-specified orbits, where these four parameters become dynamic quantities given in an additional text file.  Orbits can also be looped if desired.

The code presently allows metallicities of $Z = 0.03$--$0.0001$ with full evolutionary functionality, and a $Z=0$ mode valid only for protostellar evolution.  The latter includes opacities taken from Ref.~\refcite{Eldridge04}, but does not yet contain an implementation of the full opacity tables required to treat the case where a Pop\,III star passes from the $Z=0$ regime into the non-metal-free one by nuclear burning.  \textsf{DarkStars} comes with ZAMS starting models for all non-zero metallicities; protostellar models must be supplied by the user.

\begin{figure}[tbp]
\begin{center}
\psfig{file=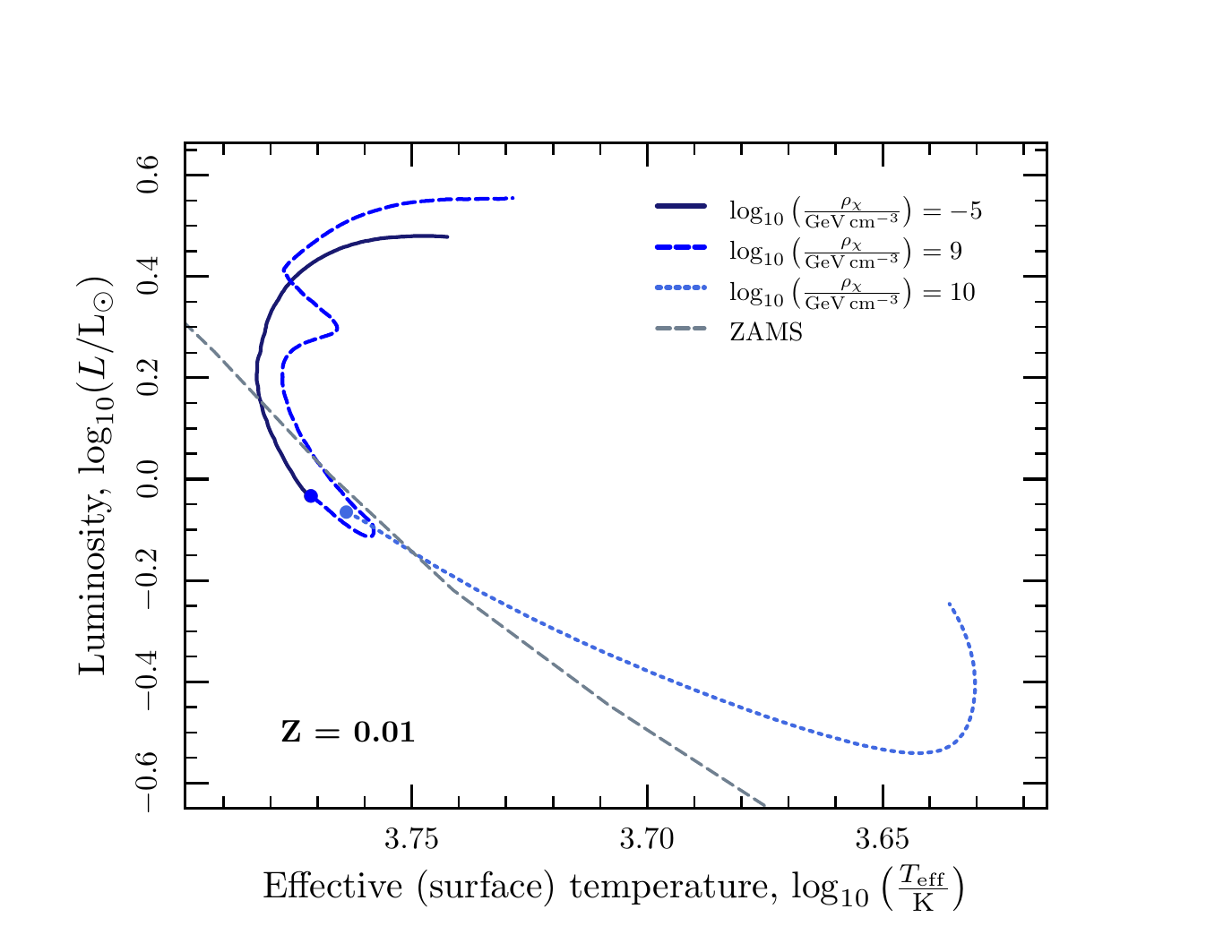,width=\textwidth}
\end{center}
\caption{Example evolutionary paths in the HR-diagram of a solar-mass star evolved in three different dark matter environments.  Filled circles indicate the approximate starting point of each track; the starting point of the star evolved in the densest halo is offset from the others only because of the large adjustment to its structure required to accommodate the effects of dark matter in converging the initial model.  In each case, stars were given a proper velocity of 220\,km\,s$^{-1}$ through their dark matter halo, which was modelled as an isothermal sphere with velocity dispersion $\sigma=270$\,km\,s$^{-1}$.  Runs were halted when the star either left the main sequence (as for the two upper curves), or ceased to move any further (the case for the lower curve).  In the case of the lower curve, WIMP annihilation provides enough energy to push the star back up the Hayashi track and hold it there, turning it into a dark star with a lifetime well beyond the current age of the Universe.}
\label{fig1}
\end{figure}

\begin{figure}[tb]
\begin{center}
\begin{minipage}{0.1\textwidth}
  \begin{center}
  \psfig{file=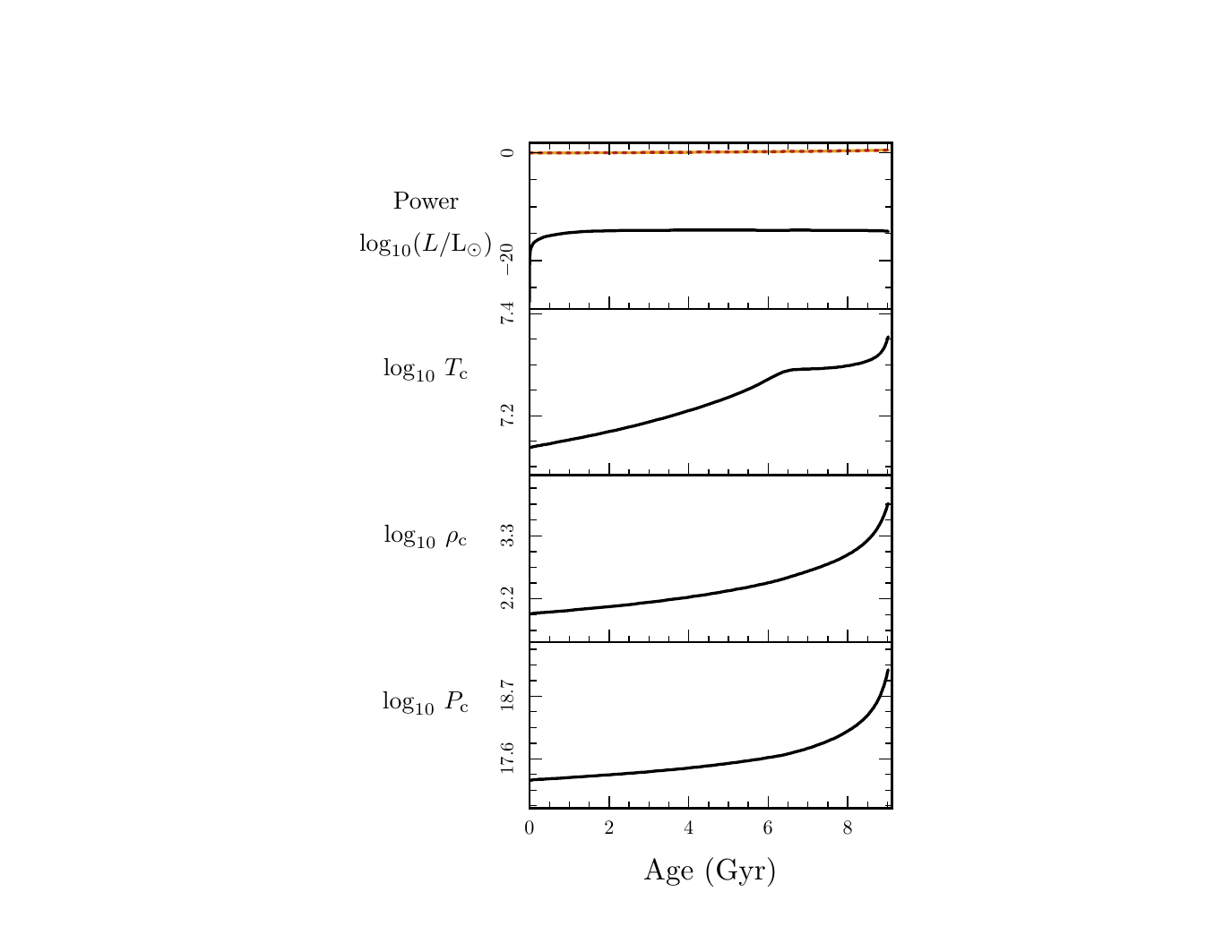, trim = 115 20 236 20, clip = true, width=\textwidth}
  \end{center}
\end{minipage}
\begin{minipage}{0.29\textwidth}
  \begin{center}
  \footnotesize{$\rho_\chi = 10^{-5}\,\mathrm{GeV}\,\mathrm{cm}^{-3}$}
  \psfig{file=ScottFig2a, trim = 160 20 100 20, clip = true, width=\textwidth}
  \end{center}
\end{minipage}
\begin{minipage}{0.29\textwidth}
  \begin{center}
  \footnotesize{$\rho_\chi = 10^{9}\,\mathrm{GeV}\,\mathrm{cm}^{-3}$}
  \psfig{file=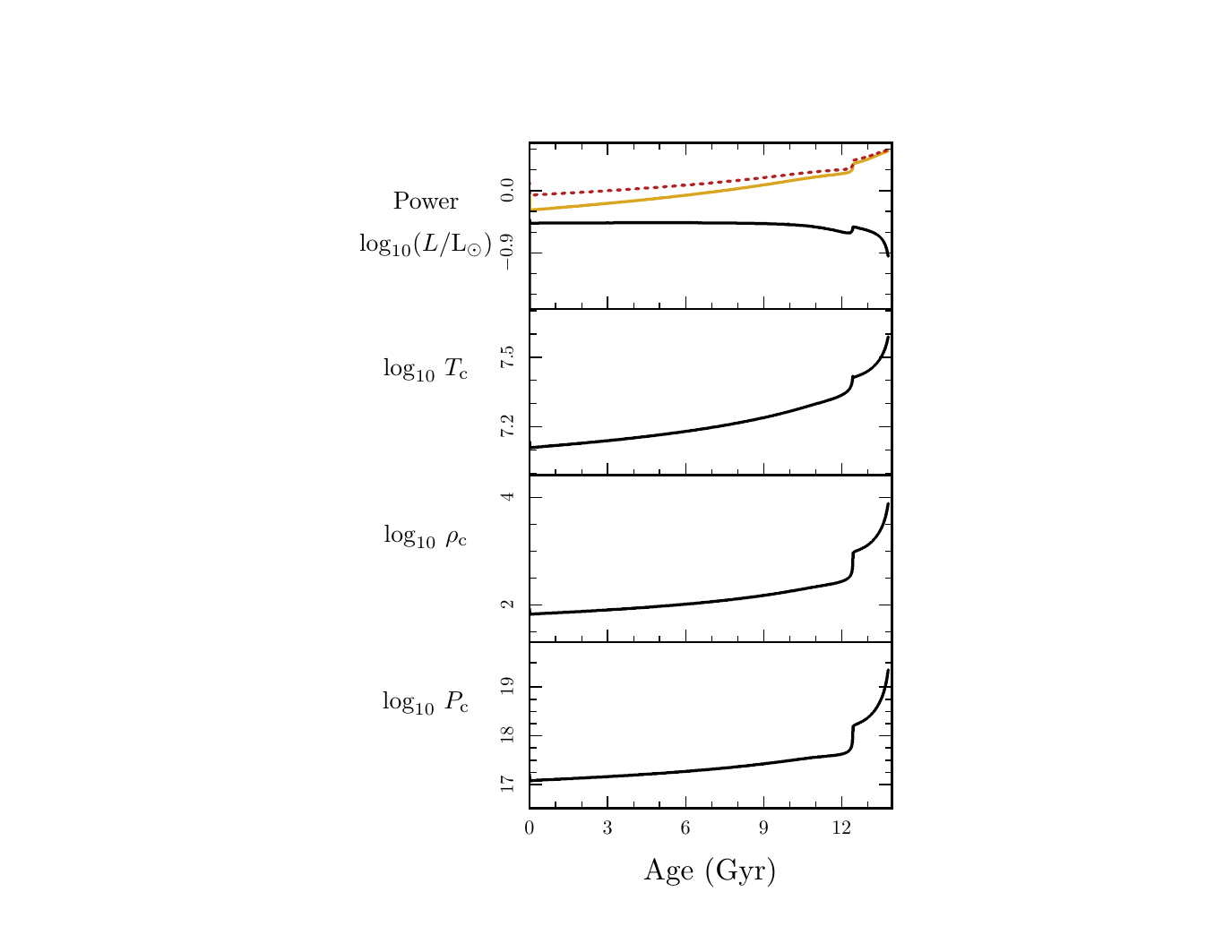, trim = 160 20 100 20, clip = true, width=\textwidth}
  \end{center}
\end{minipage}
\begin{minipage}{0.29\textwidth}
  \begin{center}
  \footnotesize{$\rho_\chi = 10^{10}\,\mathrm{GeV}\,\mathrm{cm}^{-3}$}
  \psfig{file=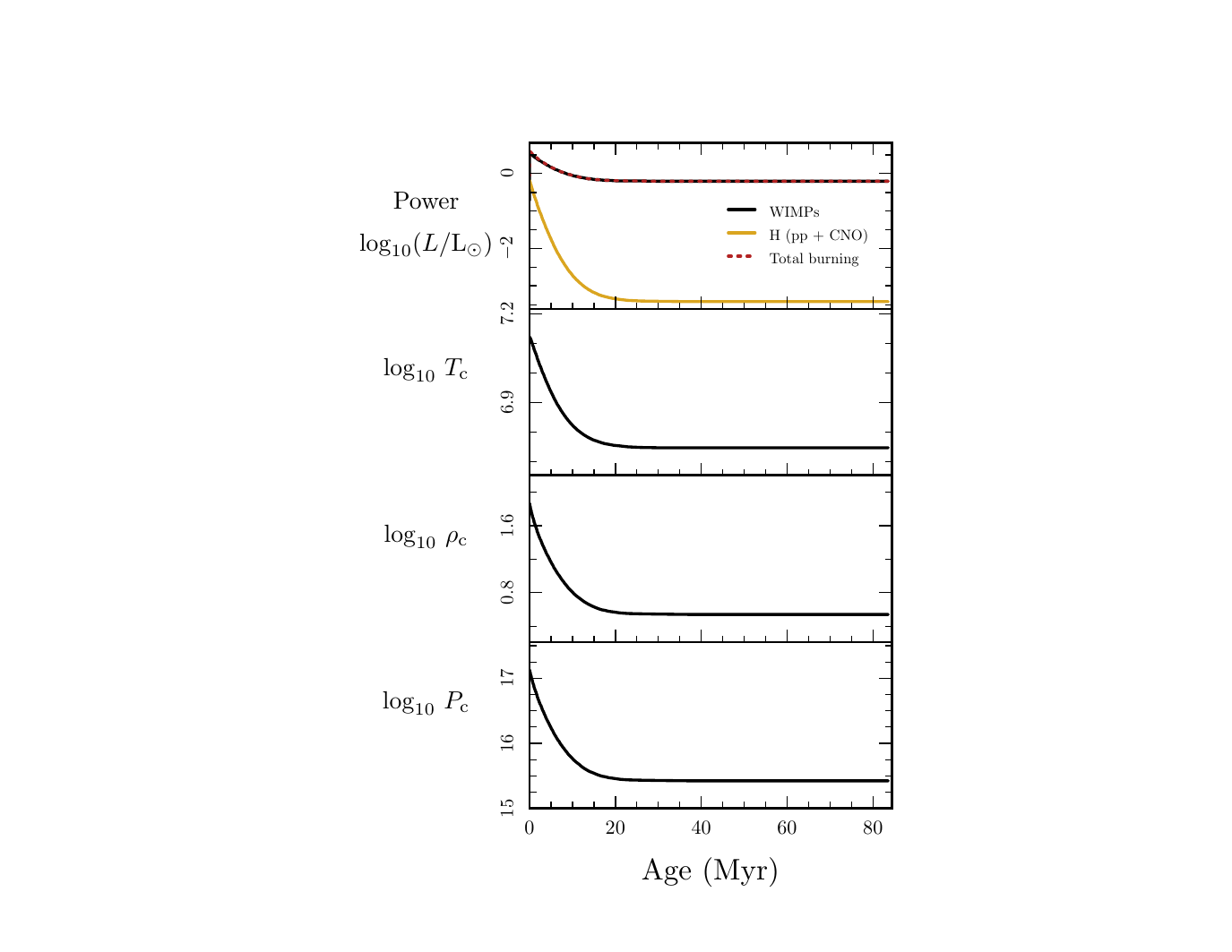, trim = 160 20 100 20, clip = true, width=\textwidth}
  \end{center}
\end{minipage}
\end{center}
\caption{Evolution of the partial luminosities provided by WIMP annihilation and fusion, as well as the central temperatures, densities and pressures of the example stars shown in Fig.~\ref{fig1}.  Note the marked drop in central temperature, density and pressure in the dark star of the rightmost panel as it regresses up the Hayashi track, drastically reducing the power provided by nuclear fusion.  Note also the extended main sequence lifetime of the middle star as compared to the normal star on the left, and its very sudden exhaustion of core hydrogen, as indicated by very steep changes in its core properties at an age of $\sim$12\,Gyr.  This sudden and violent exit from the main sequence might result in dynamical instability and/or mass ejection.}
\label{fig2}
\end{figure}

\begin{figure}[tb]
\begin{center}
\psfig{file=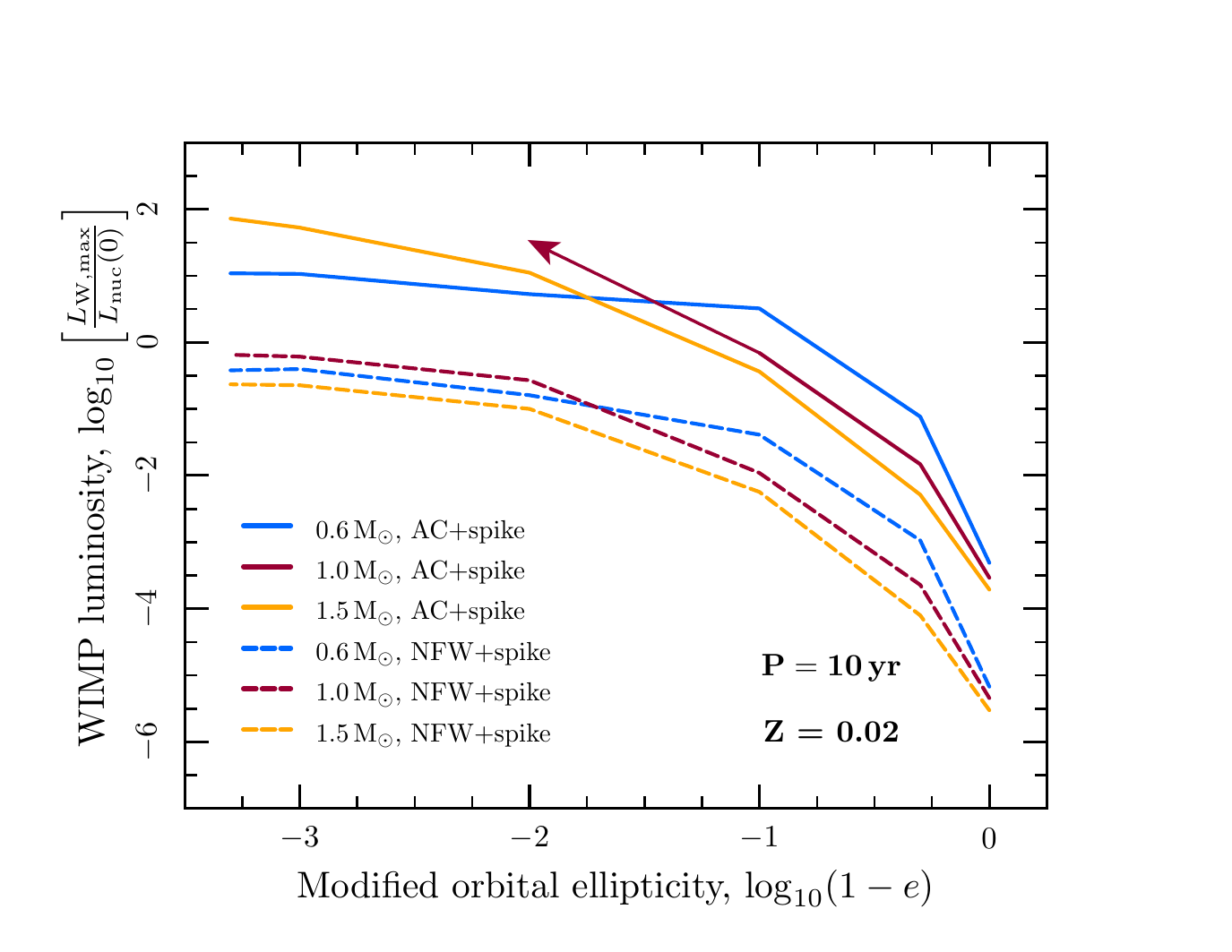,width=\textwidth}
\end{center}
\caption{WIMP-to-nuclear luminosity ratios achieved by stars on orbits with 10-year periods around the Galactic centre.  Dark matter annihilation can produce up to 100 times as much power as nuclear fusion in stars on realistic orbits in our own Galaxy.  If the Galactic halo has been adiabatically contracted (AC+spike), annihilation can equal nuclear fusion in stars on orbits with eccentricities greater than $e=0.9$, for masses less than about 1.5\,M$_\odot$.  If not (NFW+spike), stars of a solar mass or less require $e\gtrsim0.99$ to approach break-even between annihilation and fusion.  The arrow indicates that the 1\,M$_\odot$, AC+spike curve is expected to probably continue in this direction, but converging stellar models becomes rather difficult for such high WIMP luminosities.  From Ref.~\refcite{Scott09}.}
\label{fig3}
\end{figure}

In Fig.~\ref{fig1} we give some example evolutionary tracks computed with \textsf{DarkStars}, for a $Z=0.01$, $1\,\mathrm{M}_\odot$  star.  The three different paths result from immersing the star in different ambient halo densities of dark matter: one essentially without dark matter, another in a halo with a moderate dark matter density, and a third in a very dense dark matter environment.  In this case we have chosen to use a simple, non-truncated, isothermal halo velocity distribution with dispersion $\sigma=270$\,km\,s$^{-1}$, and set the star moving through it at 220\,km\,s$^{-1}$.  The runs in Fig.~\ref{fig1} have been halted when the star either leaves the main sequence, or ceases to move any further in the HR diagram.  In the most extreme case, where the ambient density is highest, the energy provided by WIMP annihilation pushes the star partially back up the Hayashi track and continues to provide enough energy to keep it there well beyond the age of the Universe; the star has become a `dark star', powered entirely by dark matter annihilation.

The interior changes producing these different evolutionary histories are summarised in Fig.~\ref{fig2}.  Here we show the changing contributions of nuclear burning and WIMP annihilation over time in the three stars from Fig.~\ref{fig1}, as well as the evolution of their central temperatures, densities and pressures.  In the case of the dark star, we see that as annihilation energy dominates over fusion power, the star's core cools and expands, driving it back up the protostellar cooling curve and effectively turning nuclear burning off.  The star evolved in the intermediately-dense dark matter halo also exhibits clear differences to the normal star, spending about 50\% longer on the main sequence.  Despite the steadily diminishing role of dark matter in this star relative to hydrogen burning over time, it also experiences a much more sudden exhaustion of its nuclear fuel than the normal star just before moving away from the main sequence, as seen in the very steep increase in its core properties just beyond 12\,Gyr.

As a taste of the sorts of more detailed studies which can be performed with \textsf{DarkStars}, in Fig.~\ref{fig3} we give an example from Ref.~\refcite{Scott09} of the sorts of WIMP-to-nuclear burning ratios that dark stars might achieve on orbits of different eccentricities, close the the centre of of the Milky Way.  All orbits in this plot had periods of 10 years.  Typically, stars in which WIMP annihilation produces more energy than nuclear fusion (i.e. above 0 on the $y$-axis of Fig.~\ref{fig3}) are considered good candidates for detection as dark stars.  This plot includes curves corresponding to two different dark matter density profiles near the Galactic centre: an adiabatically contracted halo with a gravitationally-induced spike around the black hole which has been allowed to diffuse away over time (`AD+spike'), and an NFW profile with a similar spike (`NFW+spike').  Here we see that in the adiabatic contraction scenario, WIMP annihilation can provide up to 100 times the energy of fusion for stars on realistic orbits in our own Galaxy.  The curves shown here correspond to capture from the Via Lactea-derived velocity distribution mentioned earlier, though it should be noted that the applicability of results from large-scale $N$-body simulations to velocity distributions near the central supermassive black hole is debatable; the expected distribution of WIMP velocities at the centre of the Galaxy is still quite uncertain.

\textsf{DarkStars} uses the EZ flavour\cite{Paxton04} of \textsc{stars}, making it relatively accessible to those who wish to extend or modify the code to perform a particular task.  The current version is prepared for future modifications to the WIMP velocity distributions and nuclear form factors, as well as the inclusion of WIMP evaporative effects and/or full metal-free evolution.  A legacy mode for computing capture rates by the Sun, and an (experimental) mode for computing evolution in the case where WIMPs take a non-negligible time to thermalise inside a star, are also included.

Input files for the three example evolutionary runs discussed above are included in the \textsf{DarkStars} release, along with the expected outputs from the code.  Ruby plotting scripts using Tioga\footnote{\protect\url{http://tioga.rubyforge.org/}} are also included for generating all the plots shown in these proceedings and Ref.~\refcite{Scott09}, along with the original plotting scripts included in EZ.  An option (which draws partially on the Tioga scripts originally shipped with EZ) is also provided to generate movies of a star's evolution.

\emph{Acknowledgements:} PS is grateful to the G \& E Kobbs and Helge Axelsson Johnsons Foundations for enabling his attendance at Dark2009 through their generous financial support, to Ross Church and Richard Stancliffe for helpful discussions about \textsc{stars}, and to Bill Paxton for permission to include large parts of EZ in the \textsf{DarkStars} release.  JE thanks the Swedish Research Council for funding support.

\bibliography{DMbiblio}

\end{document}